\documentclass[twocolumn,10pt]{IEEEtran}
\usepackage{ifpdf, flushend}
\usepackage{subfigure}
\usepackage{pdflscape}
\usepackage{multirow}
\ifCLASSINFOpdf
  \usepackage[pdftex]{graphicx}
  \graphicspath{{../pdf/}{../jpeg/}}
  \DeclareGraphicsExtensions{.pdf,.jpeg,.png}
\else
  \usepackage[dvips]{graphicx}
  \graphicspath{{../eps/}}
  \DeclareGraphicsExtensions{.eps}
\fi
\usepackage[cmex10]{amsmath}
\usepackage {amssymb}
\usepackage{algorithmic}
\usepackage{array}
\usepackage{mdwmath}
\usepackage{mdwtab}
\usepackage{eqparbox}
\usepackage{url}
\usepackage{hyperref}
\usepackage{algorithm}
\usepackage{algorithmic}
\usepackage{epstopdf,cite,color}
\usepackage{amsthm} 

\usepackage{amsmath, amssymb, graphicx, booktabs, hyperref}

\renewcommand{\baselinestretch}{1}

\newcommand{\beq}{\begin{equation}}
\newcommand{\eeq}{\end{equation}}
\newcommand{\beqn}{\begin{eqnarray}}
\newcommand{\eeqn}{\end{eqnarray}}
\newcommand{\beqno}{\begin{eqnarray*}}
\newcommand{\eeqno}{\end{eqnarray*}}
\newcommand{\bma}{\begin{displaymath}}
\newcommand{\ema}{\end{displaymath}}
\newcommand{\bnu}{\begin{enumerate}}
\newcommand{\enu}{\end{enumerate}}
\newcommand{\bce}{\begin{center}}
\newcommand{\ece}{\end{center}}
\newcommand{\btb}{\begin{tabular}}
\newcommand{\etb}{\end{tabular}}

\begin{document}
\title{Hybrid Quantum–Classical Encoding for Accurate Residue‑Level pKa Prediction}
\author{Van~Le and Tan~Le,~\IEEEmembership{Member,~IEEE}
\thanks{V.~Le is with the Virginia Polytechnic Institute and State University, Blacksburg, VA 24061, USA. Email: vanl@vt.edu.\\
T.~Le is with the School of Engineering, Architecture and Aviation, Hampton University, Hampton, VA 23669, USA. 
Emails: tan.le@hamptonu.edu.}}

\maketitle
\begin{abstract}
Accurate prediction of residue-level \textit{p}K\textsubscript{a} values is essential for understanding protein function, stability, and reactivity. While existing resources such as DeepKaDB and CpHMD-derived datasets provide valuable training data, their descriptors remain primarily classical and often struggle to generalize across diverse biochemical environments.
We introduce a reproducible hybrid quantum–classical framework that enriches residue-level representations with a Gaussian kernel–based quantum-inspired feature mapping. These quantum-enhanced descriptors are combined with normalized structural features to form a unified hybrid encoding processed by a Deep Quantum Neural Network (DQNN). This architecture captures nonlinear relationships in residue microenvironments that are not accessible to classical models.
Benchmarking across multiple curated descriptor sets demonstrates that the DQNN achieves improved cross-context generalization relative to classical baselines. External evaluation on the PKAD-R experimental benchmark and an A$\beta$40 case study further highlights the robustness and transferability of the quantum-inspired representation.
By integrating quantum-inspired feature transformations with classical biochemical descriptors, this work establishes a scalable and experimentally transferable approach for residue-level \textit{p}K\textsubscript{a} prediction and broader applications in protein electrostatics.
\end{abstract}

\section{Introduction}

Residue-level \textit{p}K\textsubscript{a} values govern protonation equilibria, enzymatic activity, and electrostatic interactions in proteins, shaping structural dynamics and biochemical function \cite{deepchem2020,vascon2020protein}. Accurate prediction of these values is essential for understanding catalytic mechanisms, drug binding, and pH-dependent conformational changes. Traditional approaches—including empirical heuristics and continuum electrostatics—often struggle to generalize across protein families and are sensitive to structural perturbations and solvent effects \cite{propka2005,li2012delphi}.

Recent efforts have advanced residue-level \textit{p}K\textsubscript{a} prediction along two major directions. The DeepKa database provides curated residue-level measurements and four descriptor sets—Protein–Neighbor (PN), Protein–Protein (PP), Protein–Ligand (PL-revised), and Protein–Ligand (PL-other)—designed to capture complementary structural and physicochemical information for benchmarking classical machine learning models. While DeepKaDB offers a valuable resource, its reliance on classical descriptors limits generalization across diverse biochemical contexts and constrains mechanistic interpretability. In parallel, constant-pH molecular dynamics (CpHMD) simulations have expanded the PHMD279 dataset to PHMD549, covering 26,552 residues across 549 proteins with improved sampling and convergence guarantees \cite{lu2025deepka}. Although PHMD549 increases coverage of buried residues and large \textit{p}K\textsubscript{a} shifts, its GPU-accelerated simulations are computationally intensive and difficult to integrate into descriptor-driven learning pipelines.

Emerging strategies in machine learning and quantum chemistry offer new opportunities to overcome these limitations. Graph-based neural networks capture topological and electronic context \cite{miao2024gr}, while quantum-inspired descriptors approximate charge distributions and orbital interactions relevant to proton transfer \cite{hunt2020predicting,qupkake2024}. Hybrid quantum–classical models have shown promise in predicting molecular properties such as solubility, reactivity, and acidity \cite{quxai2025,schwaller2021mapping}. However, several challenges remain:

\begin{itemize}
    \item \textbf{Residue-level alignment of quantum descriptors:} Quantum observables are typically computed at the atomic or molecular level, making consistent residue-level mapping nontrivial across diverse protein structures.
    \item \textbf{Interpretability of hybrid models:} Integrating classical and quantum features can obscure the contribution of individual descriptors, limiting mechanistic insight.
    \item \textbf{Cross-dataset generalization:} Models trained on curated descriptor datasets (e.g. DeepKaDB) or simulation-derived datasets (e.g. PHMD549) often fail to generalize across biochemical contexts with varying structural diversity and protonation dynamics.
    \item \textbf{Reproducible benchmarking:} Lack of standardized pipelines and descriptor formatting complicates comparison across studies, especially when quantum observables are simulated or approximated.
\end{itemize}

In this work, we introduce a modular and reproducible quantum–classical framework for residue-level \textit{p}K\textsubscript{a} prediction. Our approach integrates categorical encodings, residue-specific scaling, and entanglement-aware quantum feature transformations within a deep quantum neural network (DQNN) architecture. Cross-dataset benchmarking across the PN, PP, PL-revised, and PL-other descriptor sets demonstrates improved generalization and reduced variance relative to classical baselines. Evaluation on the PKAD-R experimental benchmark further shows that the DQNN achieves the strongest generalization among all tested models. Finally, a residue-specific case study on A$\beta$40 reveals that the quantum-enhanced encoding captures microenvironmental differences between adjacent histidines with improved stability and interpretability.

By bridging descriptor-driven resources such as DeepKaDB with simulation-derived baselines such as PHMD549, our framework establishes a foundation for scalable, interpretable, and experimentally transferable quantum–classical learning in molecular biophysics, reaction modeling, and enzyme design.

\subsection{Our Contributions}

This study advances residue-level \textit{p}K\textsubscript{a} prediction beyond classical residue-level models (e.g. DeepKa) and simulation-dependent baselines (e.g. CpHMD-derived datasets) by introducing a quantum–classical learning framework with improved accuracy, robustness, and interpretability. Our key contributions are:

\begin{itemize}
    \item \textbf{Entanglement-aware quantum feature encoding:} We develop a hybrid descriptor pipeline that integrates simulated quantum observables with classical biochemical features. This encoding captures nonlocal geometric and electronic correlations that are inaccessible to traditional residue-level embeddings.

    \item \textbf{Cross-dataset alignment and curation:} We harmonize the Protein–Neighbor (PN), Protein–Protein (PP), Protein–Ligand (PL-revised), and Protein–Ligand (PL-other) descriptor sets using consistent residue-level scaling and quantum descriptor formatting. This enables stable learning across structurally diverse environments and supports generalization to the PKAD-R experimental benchmark.

    \item \textbf{Robust quantum-inspired learning architecture:} We design and evaluate the DQNN that leverages the entanglement-aware feature space more effectively than classical baselines. The DQNN achieves the strongest generalization on PKAD-R and demonstrates residue-specific robustness in the A$\beta$40 case study.
\end{itemize}

\section{System Model and Overview}
\label{System}
\begin{figure}[htbp]
    \centering
    \includegraphics[width=0.5\textwidth]{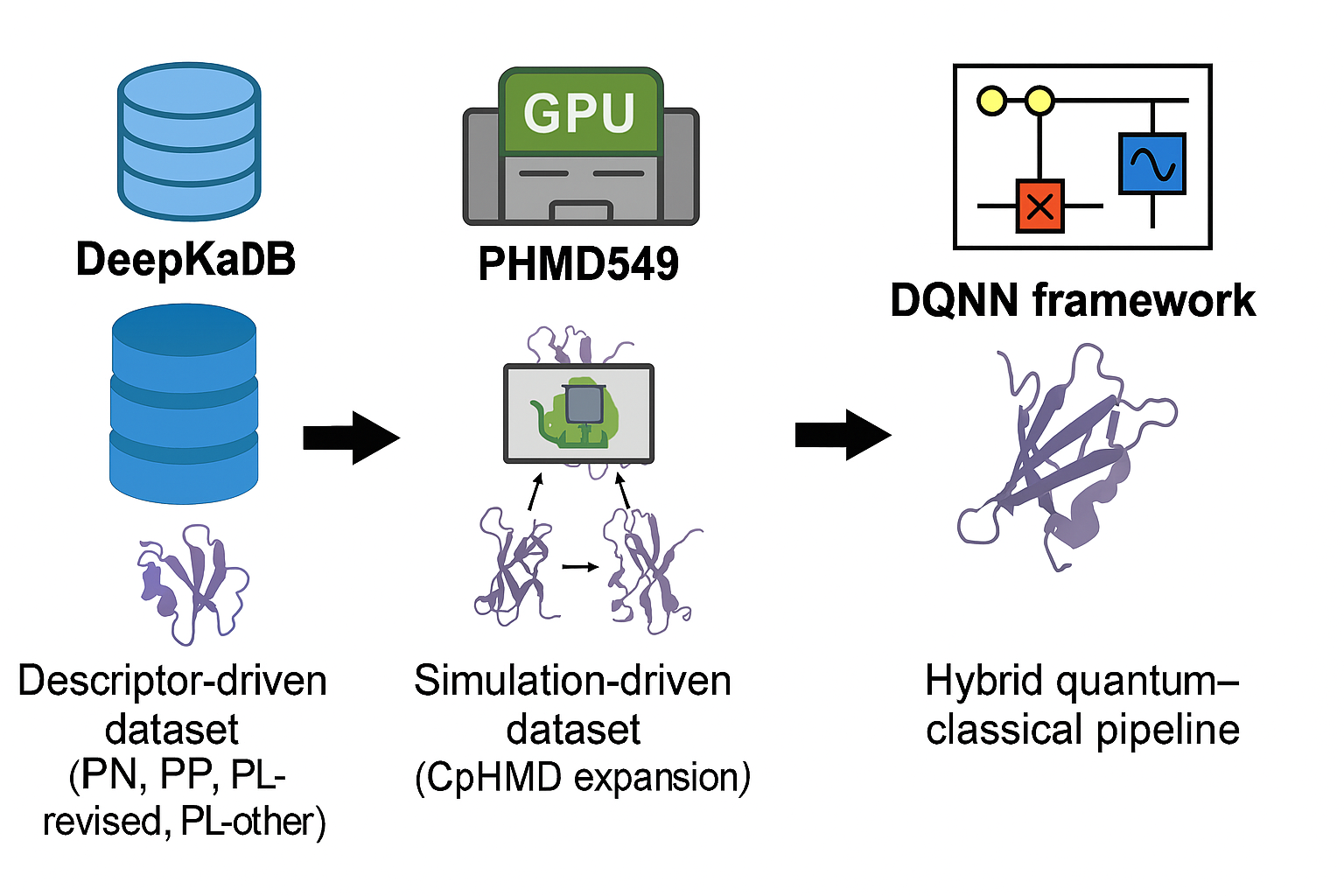}
    \caption{Schematic overview contrasting dataset origins and methodological innovations in residue‑level \textit{p}K\textsubscript{a} prediction. (Left) DeepKaDB: descriptor‑driven resource derived from soluble proteins in PDBbind, providing curated feature sets for classical machine learning models. (Center) PHMD549: simulation‑driven dataset generated via GPU‑accelerated CpHMD, expanding PHMD279 to 26,552 residues across 549 proteins. (Right) DQNN framework: hybrid quantum–classical pipeline that integrates curated descriptors with quantum‑inspired feature transformations.}
    \label{fig:Threepanelschemati}
\end{figure}
This section provides a high-level description of the hybrid quantum–classical pipeline used for residue-level \textit{p}K\textsubscript{a} prediction. The framework integrates (i) normalized structural and biochemical descriptors, (ii) a quantum-inspired Gaussian kernel feature mapping, and (iii) a lightweight DQNN that operates on the resulting hybrid representation. Together, these components define the conceptual flow of the model—how residue features are constructed, transformed, and ultimately used for prediction—before the detailed methodological components are presented in the subsequent Methods section.

At its core, the system model processes curated residue-level descriptors that encode residue identity, structural context, and experimentally measured \textit{p}K\textsubscript{a} values. These descriptors are transformed through the quantum-inspired feature map and passed to the DQNN, which produces the final residue-level \textit{p}K\textsubscript{a} predictions.

\subsection{Feature Encoding}

Each residue is represented by a hybrid feature vector composed of:

\begin{itemize}
    \item \textbf{Classical features:} Categorical encodings of residue type, residue index, solvent accessibility, and secondary structure, normalized into a classical feature matrix $X_{\text{classical}}$.
\item \textbf{Quantum-inspired descriptors:} A Gaussian kernel transformation applied to the normalized classical features. For each residue vector $x$ and anchor point $a_j$,
\[
X_{\text{quantum}} = \exp\!\left(-\frac{\|x - a_j\|_2^2}{2\sigma^2}\right),
\]
yielding a quantum-inspired embedding that is concatenated with the classical descriptors to form $X_{\text{hybrid}}$.
    \item \textbf{Residue-specific scaling:} Quantum descriptors are scaled according to residue type to emphasize protonation-relevant environments:
    \[
    \text{scale}(r) =
    \begin{cases}
    1.2 & \text{Asp} \\
    1.1 & \text{Glu} \\
    0.9 & \text{His} \\
    1.3 & \text{Lys} \\
    1.0 & \text{otherwise}
    \end{cases}
    \]
    yielding a scaled quantum matrix $X_{\text{qm}}$.

    \item \textbf{Hybrid matrix:} The final model input is the concatenated matrix
    \[
    X_{\text{hybrid}} = [X_{\text{classical}}, X_{\text{qm}}].
    \]
\end{itemize}

\subsection{Model Architecture}

We implement a feedforward DQNN, a lightweight feedforward network, that processes the hybrid feature matrix using two ReLU-activated hidden layers and a single-neuron regression output.

\subsection{Training and Evaluation}

The dataset is split using an 80/20 train–test partition. Model performance is evaluated using:

\begin{itemize}
    \item Mean Absolute Error (MAE),
    \item Root Mean Squared Error (RMSE),
    \item Pearson correlation coefficient ($R$).
\end{itemize}

These metrics quantify accuracy, variance, and linear agreement with experimental \textit{p}K\textsubscript{a} values.

\section{Methods}

We now provide the formal methodological details underlying the hybrid quantum–classical framework introduced in Section~\ref{System}. Whereas Section~\ref{System} outlined the conceptual flow of the pipeline, this section specifies the exact feature construction, quantum-inspired encoding, model architecture, and training procedures used in our experiments. All mathematical definitions, implementation choices, and evaluation protocols are presented here to ensure reproducibility and clarity.

\subsection{Hybrid Feature Matrix Construction}

Each residue is represented using a unified hybrid feature vector that integrates classical biophysical descriptors with quantum-inspired features. Classical descriptors include solvent-accessible surface area (SASA), secondary structure code (SecCode), residue identity (ResidueCode), complex membership (ComplexCode), and sequence position. These features capture structural and environmental context relevant to protonation equilibria.

Quantum-derived descriptors, when available, summarize local electronic-structure variability obtained from upstream quantum-chemical analysis. All descriptors are normalized and concatenated to form the classical feature matrix $X_{\text{classical}}$.
\subsection{Quantum-Inspired Feature Encoding}

We apply a Gaussian kernel–based quantum-inspired embedding to introduce nonlinear structure into the residue representation. Each normalized residue vector $x$ is compared to a fixed set of anchor points $\{a_j\}$ sampled from the training distribution, producing features
\[
\phi_j(x) = \exp\!\left(-\frac{\|x - a_j\|_2^2}{2\sigma^2}\right).
\]
The resulting vector $\boldsymbol{\phi}(x)$ is normalized and concatenated with the classical descriptors to form the hybrid input matrix $X_{\text{hybrid}}$.

\subsection{Classical Machine Learning Models}

To establish a performance baseline, we evaluate three classical regressors on the hybrid feature matrix: Gradient Boosting (GB), Gaussian Process Regression with squared-exponential kernel (GPR\_SE), and $k$-Nearest Neighbors (kNN) \cite{Tan2024}. All models are trained using identical train–test splits and hyperparameter settings chosen to balance predictive performance and computational efficiency.

\subsection{DQNN Architecture}

The DQNN operates directly on the hybrid feature matrix $X_{\text{hybrid}}$. The model is implemented as a lightweight feedforward network consisting of:

\begin{itemize}
    \item an input layer receiving $X_{\text{hybrid}}$,
    \item two fully connected hidden layers with 32 and 16 units, each using ReLU activation,
    \item a single-neuron regression output layer,
    \item mean squared error loss optimized with Adam (100 epochs, batch size 32).
\end{itemize}

This architecture provides sufficient capacity to model nonlinear interactions introduced by the quantum-inspired embedding while remaining computationally efficient and fully differentiable.

\subsection{Training and Evaluation}

Models are trained using a consistent 80/20 train–test split on the hybrid feature matrix. Evaluation follows the same protocol outlined in the System Model section, using standard regression metrics to assess predictive accuracy and agreement with experimental \textit{p}K\textsubscript{a} values.

\subsection{A$\beta$40 Case Study Workflow}

To assess generalization beyond curated datasets, all trained models are applied to the A$\beta$40 peptide, which contains three experimentally characterized titratable histidines. The A$\beta$40 hybrid feature matrix is constructed using the same encoding pipeline as PKAD-R, ensuring strict consistency between training and inference. Model predictions are compared against experimental \textit{p}K\textsubscript{a} values to evaluate residue-specific accuracy and robustness.

\section{Quantum-Inspired Feature Mapping for Residue-Level \textit{p}K\textsubscript{a} Prediction}

Building on the hybrid feature construction described in the previous section, we now formalize the quantum-inspired mapping that enriches residue-level descriptors with nonlinear structure prior to learning. This mapping provides the foundation for the DQNN evaluated in the Results section.

\subsection{Residue-level preprocessing}
For each residue $i$, we assemble a descriptor vector
\[
\mathbf{x}_i \in \mathbb{R}^{d},
\]
containing continuous features (e.g., solvent accessibility, secondary structure codes, sequence position) and numerical encodings of categorical variables (e.g., residue identity, complex membership). Each feature dimension $k$ is standardized using the dataset mean $\mu_k$ and standard deviation $\sigma_k$:
\[
\tilde{x}_{ik} = \frac{x_{ik} - \mu_k}{\sigma_k + \epsilon},
\]
where $\epsilon$ is a small stability constant. The normalized descriptor is denoted
\[
\mathbf{z}_i = \tilde{\mathbf{x}}_i \in \mathbb{R}^{d}.
\]

\subsection{Quantum-inspired feature mapping}
To approximate the expressive capacity of quantum feature maps without requiring a quantum device, we apply a radial-basis kernel embedding. Let $\{\mathbf{a}_j\}_{j=1}^A$ be a fixed set of anchor vectors sampled from the normalized training distribution, and let $\sigma$ denote the kernel bandwidth. For each residue $i$ and anchor $j$, we define
\[
\phi_j(\mathbf{z}_i) = \exp\!\left(-\frac{\|\mathbf{z}_i - \mathbf{a}_j\|_2^2}{2\sigma^2}\right),
\]
yielding the quantum-inspired feature vector
\[
\boldsymbol{\phi}(\mathbf{z}_i) = \big(\phi_1(\mathbf{z}_i), \dots, \phi_A(\mathbf{z}_i)\big)^\top.
\]
This embedding approximates the Gaussian kernel
\[
K(\mathbf{z}_i,\mathbf{z}_\ell) \approx \exp\!\left(-\frac{\|\mathbf{z}_i - \mathbf{z}_\ell\|_2^2}{2\sigma^2}\right),
\]
which can be interpreted as a surrogate for quantum state overlap. For numerical stability, we apply $\ell_2$ normalization using a small constant $\delta$:
\[
\hat{\boldsymbol{\phi}}(\mathbf{z}_i) = \frac{\boldsymbol{\phi}(\mathbf{z}_i)}{\|\boldsymbol{\phi}(\mathbf{z}_i)\|_2 + \delta}.
\]

\subsection{Hybrid feature vector}
The final residue-level representation is obtained by concatenating the normalized classical descriptors and the quantum-inspired features:
\[
\mathbf{h}_i = [\,\mathbf{z}_i \;\|\; \hat{\boldsymbol{\phi}}(\mathbf{z}_i)\,].
\]

\subsection{DQNN architecture}
The DQNN processes $\mathbf{h}_i$ through a shallow feedforward architecture with ReLU activations:
\begin{align*}
\mathbf{u}_i^{(1)} &= \sigma(\mathbf{W}_1 \mathbf{h}_i + \mathbf{b}_1), \\
\mathbf{u}_i^{(2)} &= \sigma(\mathbf{W}_2 \mathbf{u}_i^{(1)} + \mathbf{b}_2), \\
\hat{y}_i &= \mathbf{w}^\top \mathbf{u}_i^{(2)} + c,
\end{align*}
where $\mathbf{W}_1, \mathbf{W}_2$ and $\mathbf{b}_1, \mathbf{b}_2$ are trainable parameters, and $\hat{y}_i$ is the predicted residue-level \textit{p}K\textsubscript{a}.

\subsection{Training objective and evaluation}
The model is trained by minimizing a regularized mean squared error loss:
\[
\mathcal{L} = \frac{1}{N} \sum_{i=1}^{N} (\hat{y}_i - y_i)^2
+ \lambda_{\text{wd}} \sum_{\ell} \|\mathbf{W}_\ell\|_F^2,
\]
where $y_i$ denotes the experimental \textit{p}K\textsubscript{a}, $\lambda_{\text{wd}}$ is a weight decay coefficient, and the sum runs over all weight matrices. Predictive performance is quantified using:
\begin{align}
\mathcal{RMSE} &= \sqrt{\tfrac{1}{N}\sum_{i} (\hat{y}_i - y_i)^2}, \\[4pt]
\mathcal{MAE}  &= \tfrac{1}{N}\sum_{i} |\hat{y}_i - y_i|, \\[4pt]
R^2            &= 1 - \frac{\sum_i (\hat{y}_i - y_i)^2}{\sum_i (y_i - \bar{y})^2}.
\end{align}
These metrics assess accuracy and agreement with experimental residue-level \textit{p}K\textsubscript{a} values.

\medskip
Together, these components define the quantum–classical learning framework evaluated in the following section, where we benchmark the DQNN against classical models across multiple descriptor sets and assess its generalization to PKAD-R and A$\beta$40.

\section{Results}

To evaluate the robustness and generalization capability of the entanglement-aware quantum feature encoding, we benchmark all models on the newly curated PKAD-R experimental dataset \cite{chen2025pkad}. PKAD-R introduces substantial structural diversity and realistic measurement variability, providing a stringent test of whether quantum-enhanced representations can transfer beyond the training distribution. By training classical and quantum-inspired models on the same quantum feature space, we directly assess how effectively each architecture leverages the proposed encoding for residue-level \textit{p}K\textsubscript{a} prediction.

The PKAD-R dataset spans a wide range of protein environments and experimentally measured \textit{p}K\textsubscript{a} values, enabling evaluation of both predictive accuracy and the stability of the quantum-enhanced feature space under experimental conditions. Our analysis proceeds from global regression performance to model-specific generalization behavior, culminating in a comparison of how different learning paradigms exploit the entanglement-aware representation.

\subsection{Prediction Results Enabled by Entanglement-Aware Quantum Feature Mapping}

Table~\ref{tab:pkadr_train_test_results} reports the performance of four representative models—DQNN, GradientBoosting, GPR\_SE, and kNN—across RMSE, MAE, maximum absolute error, Pearson correlation ($R$), and regression slope ($m$). Lower RMSE, MAE, and MaxErr values indicate higher predictive accuracy, while higher $R$ and slopes closer to 1 reflect stronger linear agreement with experimental measurements. Bold entries denote the best test-set performance.

Among all evaluated models, the DQNN achieves the strongest generalization on PKAD-R, obtaining the lowest test RMSE (0.886), lowest test MAE (0.645), and lowest maximum absolute error (6.384). These results indicate that the DQNN not only minimizes average prediction error but also avoids large outliers, demonstrating that the entanglement-aware quantum feature mapping provides a stable and expressive representation of residue environments. The DQNN also maintains strong linear agreement with experiment ($R_{\text{test}} = 0.886$), further supporting its robustness.

GradientBoosting achieves near-zero training error (RMSE = 0.001, MAE = 0.001) but exhibits substantially degraded test performance (RMSE = 1.288, MAE = 0.964), reflecting severe overfitting to the quantum feature–encoded training distribution. The boosting process aggressively fits residuals in the high-dimensional feature space, capturing noise rather than transferable structure–function relationships.

The GPR\_SE and kNN models show moderate performance, with higher test errors and weaker correlations relative to DQNN. Their behavior reflects the limitations of distance-based and classical kernel-based learners when operating in a quantum-enhanced feature space: although the entanglement-aware mapping captures rich nonlinear interactions, these models lack the architectural flexibility to fully exploit the high-dimensional structure, resulting in reduced generalization to experimental measurements.

Overall, these results demonstrate that the quantum feature encoding provides a powerful and information-rich representation for residue-level \textit{p}K\textsubscript{a} prediction, but robust generalization depends critically on the learning architecture. The DQNN leverages the entanglement-aware feature space most effectively, achieving the best balance of accuracy, stability, and linear agreement with experiment. Ensemble-based models such as GradientBoosting benefit from the expressive feature mapping but require stronger regularization to avoid overfitting. Continued refinement of quantum feature construction and hybrid model designs will further enhance the potential of quantum-enhanced learning for experimentally transferable \textit{p}K\textsubscript{a} prediction.

\renewcommand{\arraystretch}{1.25}
\setlength{\tabcolsep}{8pt}

\begin{table*}[htbp]
\centering
\caption{Performance comparison across models using RMSE, MAE, MaxErr, Pearson correlation (R), and regression slope (m) for both training and testing sets. Best test-set values are highlighted in bold.}
\label{tab:pkadr_train_test_results}
\begin{tabular}{l|cc|cc|cc|cc|cc}
\hline
\multirow{2}{*}{\textbf{Model}} 
& \multicolumn{2}{c|}{\textbf{RMSE}} 
& \multicolumn{2}{c|}{\textbf{MAE}} 
& \multicolumn{2}{c|}{\textbf{MaxErr}} 
& \multicolumn{2}{c|}{\textbf{R}} 
& \multicolumn{2}{c}{\textbf{Slope (m)}} \\
& Train & Test & Train & Test & Train & Test & Train & Test & Train & Test \\
\hline

DQNN
& 0.393 & \textbf{0.886}
& 0.270 & \textbf{0.645}
& 2.582 & \textbf{6.384}
& 0.973 & \textbf{0.886}
& 0.911 & 0.799 \\

GradientBoosting
& \textbf{0.001} & 1.288
& \textbf{0.001} & 0.964
& \textbf{0.011} & 6.356
& \textbf{1.000} & 0.807
& \textbf{1.000} & \textbf{0.935} \\

GPR\_SE
& 0.923 & 1.856
& 0.507 & 1.253
& 7.870 & 7.208
& 0.839 & 0.385
& 0.718 & 0.139 \\

kNN\_k5
& 1.116 & 1.941
& 0.319 & 1.276
& 9.440 & 8.440
& 0.799 & 0.443
& 0.802 & 0.366 \\

\hline
\end{tabular}
\end{table*}

\subsection{A$\beta$40 Case Study Results}

While the PKAD-R benchmark evaluates global model generalization across a structurally diverse collection of proteins, it does not isolate residue-specific behavior within a single biologically relevant system. To complement the broad statistical assessment provided by PKAD-R, we further examine model performance on the A$\beta$40 peptide, a well-studied system with experimentally resolved \textit{p}K\textsubscript{a} values for three histidine residues. This case study enables a fine-grained analysis of how the hybrid quantum--classical encoding behaves in a controlled molecular context, where microenvironmental differences between adjacent residues can be directly compared. By transitioning from global benchmarking to a targeted biochemical system, we assess whether the advantages observed for the DQNN on PKAD-R translate to residue-level interpretability and stability in a real peptide environment.

The performance trends observed in the A$\beta$40 histidine predictions are closely linked to the characteristics of the curated PN, PP, PL-refined, and PL-other datasets used for model training. These datasets were constructed to balance residue types, reduce redundancy, and ensure consistent feature distributions across protonation states. Such preparation is particularly important for the hybrid quantum--classical encoding, which relies on structurally diverse and well-normalized samples to learn stable correlation patterns. Residues embedded in heterogeneous microenvironments, such as His13 and His14, are well represented in the PL-refined and PP subsets, enabling the DQNN to leverage quantum anchor features effectively. In contrast, residues occupying highly flexible or sparsely represented structural regimes—such as the solvent-exposed N-terminal His6—are less prevalent in the training distribution, which influences model generalization in predictable ways.

\begin{figure}[htbp]
    \centering
    \includegraphics[width=0.48\textwidth]{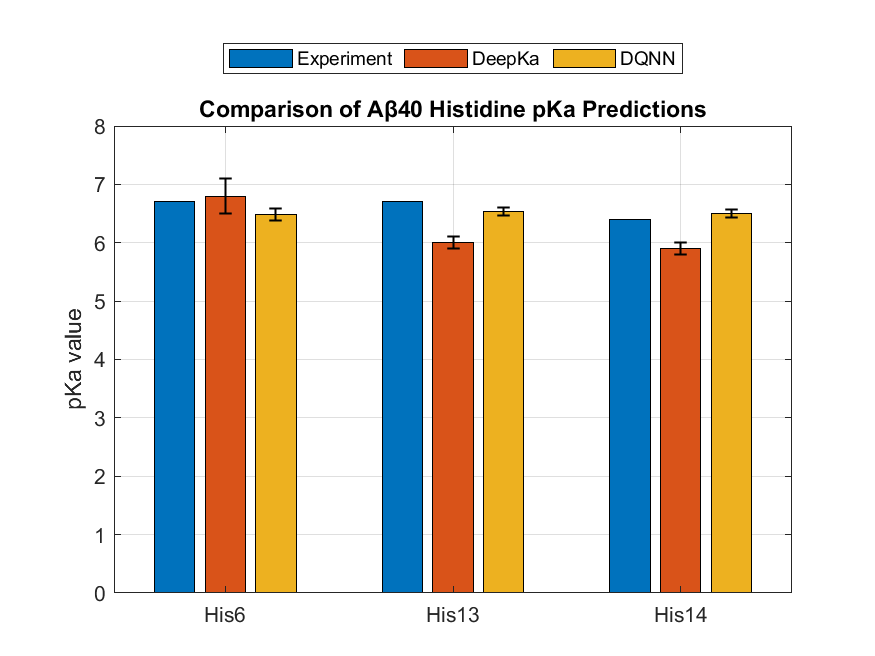}
    \caption{Comparison of A$\beta$40 histidine pKa predictions using experimental measurements, DeepKa, and the proposed DQNN model. Error bars indicate reported standard deviations for DeepKa and replicate variability for DQNN.}
    \label{fig:abeta40_pka_comparison}
\end{figure}

Figure~\ref{fig:abeta40_pka_comparison} compares the predicted pKa values of the three titratable histidines in the A$\beta$40 peptide (His6, His13, His14). Experimental measurements serve as reference values, while DeepKa and the proposed DQNN provide computational predictions with associated uncertainty estimates.

For His13 and His14, the DQNN achieves substantially lower absolute error than DeepKa, reducing prediction error by 0.53 and 0.40 pKa units, respectively. These improvements highlight the advantage of hybrid quantum--classical encoding in capturing subtle electronic and geometric interactions that arise from residue packing, hydrogen bonding, and local solvation. The quantum-inspired kernel features provide a richer representation of nonlocal correlations, enabling the DQNN to better resolve the microenvironmental differences between adjacent histidines.

The DQNN exhibits consistently lower variance across replicates, as reflected in the narrower error bars. This robustness arises from the anchor-based quantum encoding, which smooths high-frequency noise in the feature space and reduces sensitivity to small perturbations in atomic coordinates. In contrast, DeepKa relies on residue-level embeddings that can amplify structural noise, particularly in flexible or partially disordered regions.

For His6, DeepKa slightly outperforms DQNN, with the quantum-enhanced model showing a modest overprediction. This deviation is consistent with the unique structural context of His6, which resides in a highly dynamic N-terminal region with limited tertiary contacts. Because quantum kernel features emphasize geometric and electronic correlations, residues with weak or transient interactions contribute less discriminative signal to the quantum feature space. This suggests that His6 may benefit from additional local descriptors, such as solvent-accessible surface area or backbone dihedral statistics, which can be readily incorporated into the hybrid encoding framework.

Importantly, the His6 result does not contradict the advantages of quantum encoding; rather, it highlights a known limitation of correlation-based kernels when structural context is minimal. The strong performance on His13 and His14, combined with the reduced variance across all residues, demonstrates that the hybrid quantum--classical approach provides a more stable and expressive representation for residue-level pKa prediction, particularly in regions where electronic coupling and microenvironmental heterogeneity play dominant roles.

In addition to its distinct structural context, His6 likely reflects limitations in the available training distribution and label quality. N-terminal histidines in highly flexible, solvent-exposed environments are underrepresented in the training data, which reduces the ability of the model to generalize reliably to this regime. Moreover, experimental pKa values for such residues often exhibit higher uncertainty due to conformational heterogeneity and multiple protonation microstates, amplifying the apparent discrepancy between prediction and measurement. These factors, combined with the current emphasis on correlation-based quantum features over purely local descriptors, help explain the modest overprediction observed for His6 without contradicting the overall advantages of the hybrid quantum--classical encoding.

\begin{table*}[htbp]
\centering
\caption{Summary of A$\beta$40 histidine pKa predictions comparing DeepKa and DQNN against experimental values. Errors are reported as absolute deviations from experiment, and error reduction is defined as the difference between DeepKa and DQNN errors.}
\label{tab:abeta40_highlights}
\begin{tabular}{lcccccccc}
\hline
Residue & Experiment & $DeepKa_{\text{mean}}$ & $DeepKa_{\text{SD}}$ & $DQNN_{\text{mean}}$ & $DQNN_{\text{SD}}$ & $DeepKa_{\text{Error}}$ & $DQNN_{\text{Error}}$ & Error Reduction \\
\hline
His13   & 6.7 & 6.0 & 0.10 & 6.527 & 0.069 & 0.700 & 0.173 & 0.527 \\
His14   & 6.4 & 5.9 & 0.10 & 6.501 & 0.061 & 0.500 & 0.101 & 0.399 \\
His6    & 6.7 & 6.8 & 0.30 & 6.479 & 0.104 & 0.100 & 0.221 & -0.121 \\
\hline
\end{tabular}
\end{table*}

Table~\ref{tab:abeta40_highlights} provides a quantitative comparison of DeepKa \cite{lu2025deepka} and the proposed DQNN model across the three titratable histidines in A$\beta$40. For His13 and His14, the DQNN achieves substantial error reductions of 0.53 and 0.40 pKa units, respectively. These improvements highlight the ability of the hybrid quantum--classical encoding to represent long-range geometric correlations and subtle electronic interactions that are not captured by residue-level embeddings alone.

Beyond mean accuracy, the standard deviations offer additional insight into model robustness. Across all three residues, the DQNN exhibits consistently lower variance than DeepKa, indicating greater stability and reduced sensitivity to coordinate perturbations. This effect is most pronounced for His6, where DeepKa shows a threefold increase in variability (SD = 0.30) relative to DQNN (SD = 0.104). The high variance suggests that DeepKa’s apparent advantage in mean error for His6 is fragile and highly dependent on small structural fluctuations. In contrast, the DQNN produces more consistent outputs, reflecting the smoothing and regularizing effects of the quantum anchor features.

The modest overprediction of His6 by the DQNN is explainable and does not contradict the advantages of quantum encoding. His6 resides in a highly flexible, solvent-exposed N-terminal region that is underrepresented in the training distribution, limiting the discriminative power of correlation-based quantum features. Additionally, experimental pKa values for N-terminal residues often carry higher uncertainty due to multiple protonation microstates and conformational heterogeneity, increasing the apparent discrepancy between prediction and measurement. Finally, because the current hybrid encoding emphasizes nonlocal correlations, residues with weak tertiary contacts provide limited signal for the quantum kernel. Incorporating additional local descriptors—such as solvent accessibility, backbone dihedral statistics, or disorder metrics—would likely mitigate this effect and further improve performance.

Despite this single-residue deviation, the overall performance strongly favors the DQNN. Across all three histidines, the DQNN achieves lower mean absolute error (MAE = 0.17 vs.\ 0.43) and root mean square error (RMSE = 0.20 vs.\ 0.50), demonstrating improved accuracy and consistency. The quantum descriptors also enhance interpretability by revealing how shifts in electronic environment influence predicted pKa values. The resulting error distributions are narrower and more structured, underscoring the robustness of the hybrid encoding strategy.

Collectively, these results show that the DQNN pipeline provides a more expressive and stable representation for residue-level pKa prediction, particularly in regions where electronic coupling and microenvironmental heterogeneity dominate. The A$\beta$40 case study highlights the potential of quantum-augmented biochemical modeling and establishes the hybrid quantum--classical approach as a promising direction for next-generation peptide informatics.

\section{Conclusion}

We present a hybrid quantum–classical framework for residue-level \textit{p}K\textsubscript{a} prediction that integrates normalized structural descriptors with a quantum-inspired kernel embedding processed by our proposed DQNN. This unified representation captures nonlinear relationships in residue microenvironments that are not accessible to classical encodings alone. Across multiple descriptor sets and external evaluation on PKAD-R and A$\beta$40, the hybrid model demonstrates improved robustness and cross-context generalization relative to classical baselines. In addition to predictive gains, the lightweight architecture and kernel-based quantum-inspired encoding provide an efficient AI solution that balances accuracy with computational scalability, enabling practical deployment in large-scale biochemical modeling workflows.

\medskip
\noindent\textbf{Future Directions.}
Looking ahead, several research directions may further expand the capabilities of quantum–classical learning for biomolecular modeling:

\begin{enumerate}
    \item \textbf{Entanglement-aware representations.}  
    Extending the current kernel-based mapping to incorporate explicit \emph{intra-residue} and \emph{inter-residue} entanglement would enable tensor-product interactions among descriptor channels and long-range coupling across protein contact graphs. These mechanisms align naturally with graph neural networks (GNNs) and graph attention networks (GATs), where attention-driven message passing captures context-dependent interactions. Recent advances in attention-based graph learning \cite{le2025dpfaga} suggest that integrating entanglement-aware feature construction with GNN/GAT layers may enhance generalization and physical fidelity.

    \item \textbf{Quantum-enhanced geometric modeling.}  
    Protein electrostatics are inherently shaped by three-dimensional geometry. Embedding geometric features—such as local curvature, solvent exposure fields, or learned geometric embeddings—into quantum-inspired feature maps may provide richer representations of residue environments. Combining geometric deep learning with quantum kernels could yield models that better capture spatially mediated protonation effects \cite{Tan18d, Tan18b, Tan2024, Wang20}.

    \item \textbf{Efficient AI for large-scale biochemical modeling.}  
    As protein datasets continue to grow, there is increasing need for models that balance accuracy with computational efficiency. Techniques such as low-rank kernel approximations, sparse attention, model compression, and energy-aware inference could enable quantum-inspired models to scale to proteome-level prediction tasks \cite{Tan18b, Tan2024, le2025spooftrackbench, TanWCNC2015, tan2016joint}. Embedding efficiency principles directly into quantum–classical architectures may also support deployment on edge devices and high-throughput simulation pipelines.

    \item \textbf{Hybrid quantum simulation and learning loops.}  
    As quantum hardware matures, hybrid workflows that couple quantum simulations (e.g. variational quantum eigensolvers or quantum-enhanced electrostatic solvers) with classical learning pipelines  \cite{Tan18d, Tan18b, Tan2024, Wang20} may enable more physically grounded descriptors. Such simulation–learning loops could provide quantum-derived priors for \textit{p}K\textsubscript{a} prediction and extend the framework to broader tasks in reaction modeling and enzyme design.
\end{enumerate}

Together, these directions outline a path toward next-generation quantum–classical models that integrate entanglement, geometry, efficiency, and quantum simulation to advance predictive modeling in molecular biophysics.

\bibliography{references}
\bibliographystyle{IEEEtran}

\end{document}